\documentclass[12pt]{article}
\usepackage{scicite}
\usepackage{times}
\usepackage{placeins}
\usepackage{caption}
\usepackage{subcaption}
\usepackage{amsmath}
\usepackage{gensymb}
\usepackage{graphicx}
\usepackage{xcolor}
\usepackage{titling}
\usepackage{url}
\usepackage[labelfont=bf]{caption}
\makeatletter
\renewcommand{\fnum@figure}{Fig. \thefigure}

\usepackage{xr}

\makeatother
\newenvironment{sciabstract}{%
\begin{quote} \bf}
{\end{quote}}
\title{Stabilizing Magnetic Bubble Domains in Epitaxial 2D Magnet/Topological Insulator Heterostructures through Interfacial Interactions}
\author
{Thow Min Jerald Cham$^{1,2\dag\ast}$, Mowen Zhao$^{3\dag}$, Wenyi Zhou$^{4\dag}$, Andrew Koerner$^{3}$,\\ Dang-Khoa Le$^{5}$, Ziling Li$^{4}$, Lukas Powalla$^{6}$, Derek Bergner$^{3}$, \\ Eklavya Thareja$^{5}$, Camelia Selcu$^{4}$, Sadikul Alam$^{7}$, Sebastian Wintz$^{8}$,\\ Markus Weigand$^{8}$, Jinwoo Hwang$^{7}$, Jacob Gayles$^{5}$,\\ Roland Kawakami$^{4}$ and Yunqiu Kelly Luo$^{1,3,9,10,11\ast}$\\
\normalsize{}\\
\normalsize{Affiliations: $^{1}$Department of Physics, Cornell University, Ithaca, NY, USA}\\
\normalsize{$^{2}$Department of Physics, California Institute of Technology, Pasadena, CA, USA}\\
\normalsize{$^{3}$Department of Physics and Astronomy, University of Southern California,}\\
\normalsize{Los Angeles, CA, USA}\\
\normalsize{$^{4}$Department of Physics, The Ohio State University, Columbus, OH, USA}\\
\normalsize{$^{5}$Department of Physics, University of South Florida, Tampa, FL, USA}\\
\normalsize{$^{6}$ Max Planck Institute for Solid State Research, Stuttgart, Germany}\\
\normalsize{$^{7}$Department of Materials Science and Engineering, The Ohio State University,}\\
\normalsize{Columbus, OH, USA}\\
\normalsize{$^{8}$ Institut für Nanospektroskopie, Helmholtz-Zentrum Berlin für Materialien und Energie,}\\
\normalsize{Berlin, Germany}\\
\normalsize{$^{9}$Kavli Institute at Cornell, Ithaca, NY USA}\\
\normalsize{$^{10}$Mork Family Department of Chemical Engineering and Materials Science,}\\ 
\normalsize{University of Southern California, Los Angeles, CA, USA}\\
\normalsize{$^{11}$Department of Chemistry, University of Southern California, Los Angeles, CA, USA}\\
\normalsize{$^{\ast}$To whom correspondence should be addressed:}\\ \normalsize{kelly.y.luo@usc.edu, tmjcham@caltech.edu}\\
\normalsize{$^{\dag}$These authors contributed equally}
}

\date{}
\begin{document} 
\enlargethispage{3\baselineskip}
\maketitle 
\baselineskip=24pt

\newpage
\begin{sciabstract}
Epitaxial heterostructures of two-dimensional (2D) van der Waals (vdW) magnets and topological insulators offer a powerful platform for probing interfacial spin interactions that govern magnetic textures in low-dimensional quantum systems, while simultaneously enabling highly efficient, atomically thin spin–orbit-torque (SOT) memory and computing architectures.  Despite this promise, the fundamental role of these interfacial interactions in determining magnetic domain-phase stability remain largely uncharted. Here, we perform scanning transmission X-ray microscopy to image nanoscale magnetic textures in epitaxial Fe$_3$GeTe$_2$/Bi$_2$Te$_3$ heterostructures, enabled by a thermal-release-tape dry-transfer process onto X-ray transparent silicon-rich nitride membranes. Under zero-field-cooled (ZFC) conditions, we observe robust bubble-domain phases from 75 – 165 K, and across different number of folds of the multilayer [FGT/ Bi$_2$Te$_3$]$_n$ heterostructure (n = 1 to 5). This is in stark contrast with exfoliated single-crystal Fe$_3$GeTe$_2$ flakes, where ZFC stripe domains are observed for flakes thicker than 20 nm and no domains have been reported for thin flakes less than 15 nm. First-principles calculations and micromagnetic simulations reveal that interfacial coupling to Bi$_2$Te$_3$ modifies the magnetic anisotropy and introduces interfacial Dzyaloshinskii–Moriya interaction, shifting the magnetic phase space towards bubble-domain stabilization without field-cooling. Together, our results offer a new strategy for phase-selective control of magnetic domains through interfacial engineering.
\end{sciabstract}

\subsection*{Introduction}
Van der Waals (vdW) magnets have attracted significant interest for their robust magnetic properties persisting down to atomically thin layers, offering a versatile platform for probing fundamental 2D magnetism and advancing spintronics applications \cite{geim2013, huang2017layer, Deng2018, Fei2018, jiang2018screening, lee2021magnetic}. Their universal layer-by-layer compatibility further enables the exploration of rich interfacial phenomena when integrated with other vdW layers, including spin–orbit torques (SOT) \cite{Alghamdi2019-ip, Wang_2020_FGTPt, Gupta2020}, proximity-induced anisotropy and Curie temperature enhancement driven by interfacial spin–orbit coupling \cite{Zhou2023-is, Fanchiang2018_YIG, Liu2020_YIG, Wu_2021_CoFe}, and exchange coupling \cite{Zhu2020_FGTCrCl3, Zhang2022_FGTCrOCl, cham2024exchange, Balan2024_FGTMnPS3}. Recent demonstrations of high SOT efficiencies in vdW magnet/topological insulator (TI) heterostructures highlight the promise of these systems for ultralow-power magnetic memory and logic \cite{Mellnik2014-ev, wang2015topological, liu2021temperature, He2022-lx, han2023coherent, Zhou2023-is, wang2024field}. However, despite growing interest in these heterostructures, how the TI layer can play a role in altering the domain morphology through the balance between magnetic anisotropy, exchange stiffness and interfacial Dzyaloshinskii–Moriya interaction(DMI), remains largely unexplored. This knowledge gap is especially notable given that Fe$_3$GeTe$_2$ exhibits a rich set of field- and history-dependent domain phases, suggesting that even subtle interfacial perturbations could dramatically reshape its magnetic texture.

\vspace{0.5cm}
In this study, we focus on investigation of the nanoscale magnetic domain behavoir in vdW ferromagnet/vdW TI heterostructures composed of Fe$_3$GeTe$_2$ (FGT) and Bi$_2$Te$_3$. FGT is a vdW metallic ferromagnet with a Curie temperature near 200 K. Flakes exfoliated from bulk single crystals host a wide variety of domain phases whose stability depends on thickness, temperature, and magnetic history \cite{WangHong2020, Ding2020, Chakraborty2022, Birch2022-oh, Powalla2023_skyrmionium, Birch_2024, Sun2025_FGTStrain, Garland2025}. Recent XMCD and Lorentz-TEM studies mapping the magnetic phase diagram of exfoliated flakes show that ZFC conditions produce stripe domains, with bubble or skyrmion phases arising only under applied fields or field-cooling. Furthermore, these bubble or skyrmion phases are only observed in exfoliated FGT flakes thicker than 20 - 30 nm, while no domains are seen in layers thinner than 15 nm. \cite{WangHong2020, Birch2022-oh, Garland2025} Interfacing FGT flakes with layered materials such as WTe$_2$ \cite{Wu2020-yy}, Cr$_2$Ge$_2$Te$_6$\cite{Wu2022-yy}, or (Co/Pd)$_n$ \cite{Yang2020_CoPd} superlattices has been shown to introduce interfacial DMI that stabilize chiral spin textures. Yet proximity effects from topological insulators, where the strong spin–orbit coupling can simultaneously modify magnetic anisotropy \cite{Fanchiang2018, Liu2020TSS, Ye2025} and induce chiral interactions on the equilibrium magnetic domain morphology of vdW ferromagnets, remains largely unexplored. This gap has persisted largely because direct, nanoscale imaging of magnetic textures in full epitaxial vdW magnet/TI heterostructures has not previously been feasible.

\vspace{0.5cm}
Here, we overcome this challenge by combining large-area molecular beam epitaxy (MBE) with a thermal-release-tape membrane transfer technique, enabling direct scanning transmission X-ray microscopy (STXM) imaging of magnetic textures in fully epitaxial FGT/Bi$_2$Te$_3$ heterostructure \cite{Zhou2023-is,li2025fullfilmdrytransfermbegrown}. This capability reveals that proximity to the topological insulator fundamentally reshapes the magnetic domain-phase landscape of FGT, stabilizing a robust bubble-domain regime under zero-field-cooled conditions across a wide range of temperatures and effective FGT thicknesses. By integrating these measurements with first-principles density functional theory calculations and micromagnetic simulations, we show that interfacial coupling with the TI introduces an interfacial DMI, and through altering the competition between anisotropy, exchange and DMI, shifts the system into a bubble-stabilized region of magnetic domain phase space. Together, these results establish epitaxial FGT/Bi$_2$Te$_3$ heterostructures as a model platform for engineering proximity-driven magnetic textures in low-dimensional magnetic heterostructures. 

\subsection*{Magnetic Characteristics of Epitaxial Fe$_3$GeTe$_2$/Bi$_2$Te$_3$ Heterostructures}
\begin{figure}[htpb]
    \centering
    \includegraphics[width=\textwidth]{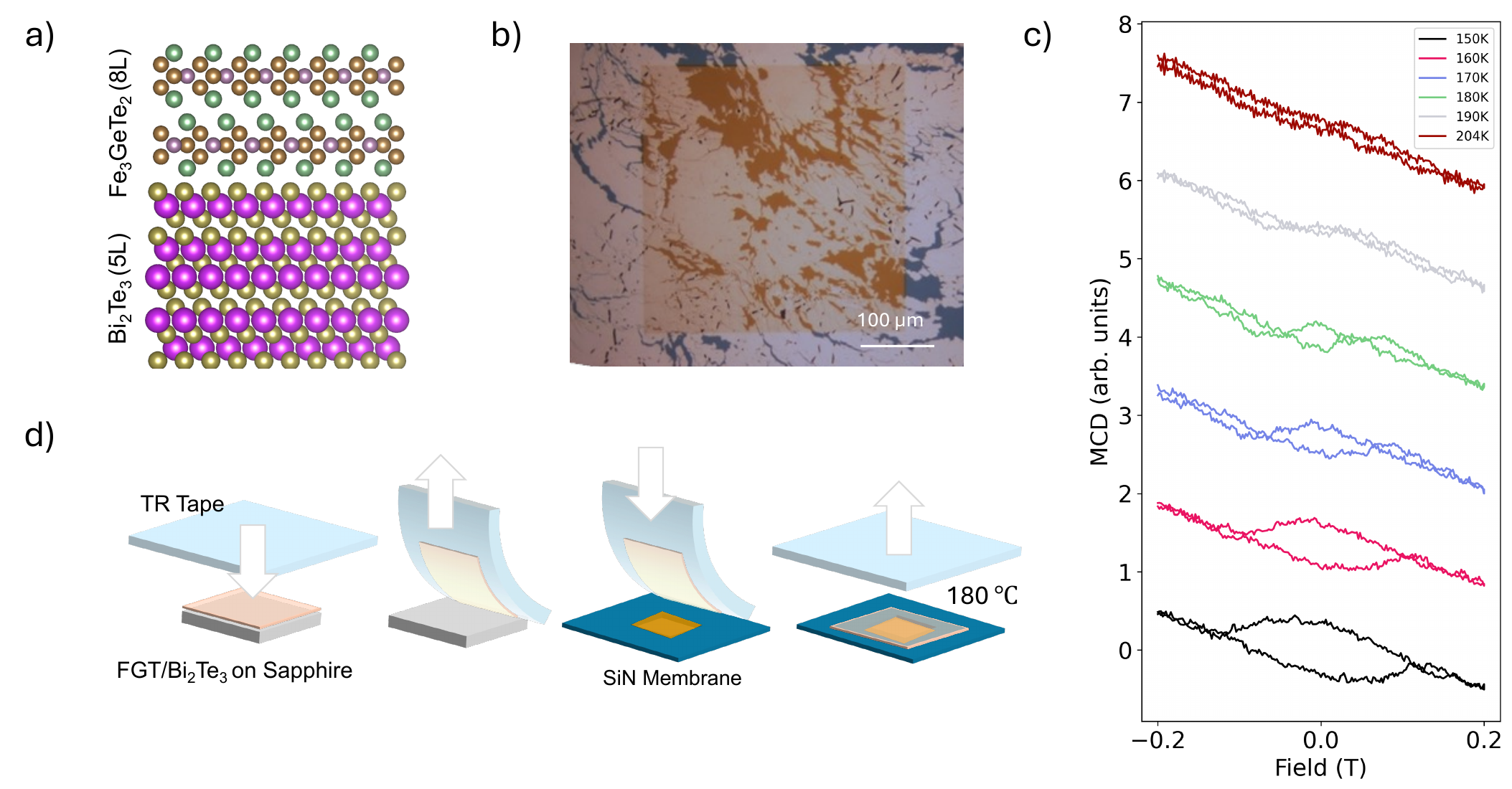}
    \caption{\textbf{MBE-grown epitaxial Fe$_3$GeTe$_2$/Bi$_2$Te$_3$ heterostructures. } (a) Schematic of atomic structure of the van der Waals magnet/topological insulator heterostructure. consisting of five quintuple layers (QL) of Bi$_2$Te$_3$ and eight van der Waals layers of Fe$_3$GeTe$_2$ (partial layers shown for clarity). (b) Optical micrograph of a large-area Fe$_3$GeTe$_2$/Bi$_2$Te$_3$ film transferred onto an X-ray--transparent silicon-rich nitride (SiRN) membrane using a thermal-release-tape (TRT) process (scale bar: 100~$\mu$m). (c) Optical magnetic circular dichroism (MCD) hysteresis loops measured at different temperatures, demonstrating strong perpendicular magnetic anisotropy and a Curie temperature of approximately 190~K. (d) Schematic illustration of the full-film TRT transfer process, enabling the transfer of epitaxial Fe$_3$GeTe$_2$/Bi$_2$Te$_3$ heterostructures from sapphire substrates onto SiRN membranes for XMCD-STXM measurements.}
    \label{Figure1}
\end{figure}

Using molecular beam epitaxy (MBE), Bi$_2$Te$_3$ thin films were first synthesized on Al$_2$O$_3$ (0001) via a two-step growth protocol, followed by epitaxial deposition of FGT (S.I.I). Continuous in-situ RHEED monitoring during the growth of both layers ensures layer-by-layer growth and accurate thickness control (Fig.~S1), as we have previously demonstrated \cite{Zhou2023-is,Goff2024-uv}. The chemical compatibility of the two compounds enables the formation of atomically sharp, interdiffusion-free interfaces. For this study, we choose a heterostructure consisting of five quintuple layers of Bi$_2$Te$_3$ ($\approx$ 5.3 nm) and eight vdW layers of FGT ($\approx$ 6.4 nm) to optimize magnetic contrast at the Fe L$_3$-edge for STXM imaging. The entire heterostructure was capped \emph{in-situ} with 10 nm Te prior to air exposure to prevent oxidation.

\vspace{0.5cm}
We first characterize the macroscopic magnetic response of the as-grown FGT/Bi$_2$Te$_3$ films using optical reflective magnetic circular dichroism (MCD) in a variable-temperature closed-cycle magneto-optical cryostat (S.I.II). The MCD hysteresis loops reveal strong perpendicular magnetic anisotropy, with square out-of-plane hysteresis and nearly 100 \% remanence below the Curie temperature, confirming robust ferromagnetism with $T_c \approx 190 $--$ 200$ K (Fig.~\ref{Figure1}c). Importantly, X-ray magnetic circular dichroism based on scanning transmission X-ray microscopy (XMCD-STXM) requires the films to be transferred onto X-ray–transparent silicon-rich nitride (SiRN) membranes (Fig.~\ref{Figure1}b). To enable this for MBE-grown heterostructures, we employ a full-film thermal release tape (TRT) transfer process (Fig.~\ref{Figure1}d), recently demonstrated to preserve epitaxy and interface integrity during membrane transfer \cite{li2025fullfilmdrytransfermbegrown} (see details in S.I.II). This workflow enables direct nanoscale XMCD-STXM characterization of epitaxial FGT/Bi$_2$Te$_3$ heterostructures and forms the foundation for the domain imaging results presented below. 

\subsection*{XMCD-STXM imaging of magnetic domains}
\begin{figure}[htpb]
    \centering
    \includegraphics[width=\textwidth]{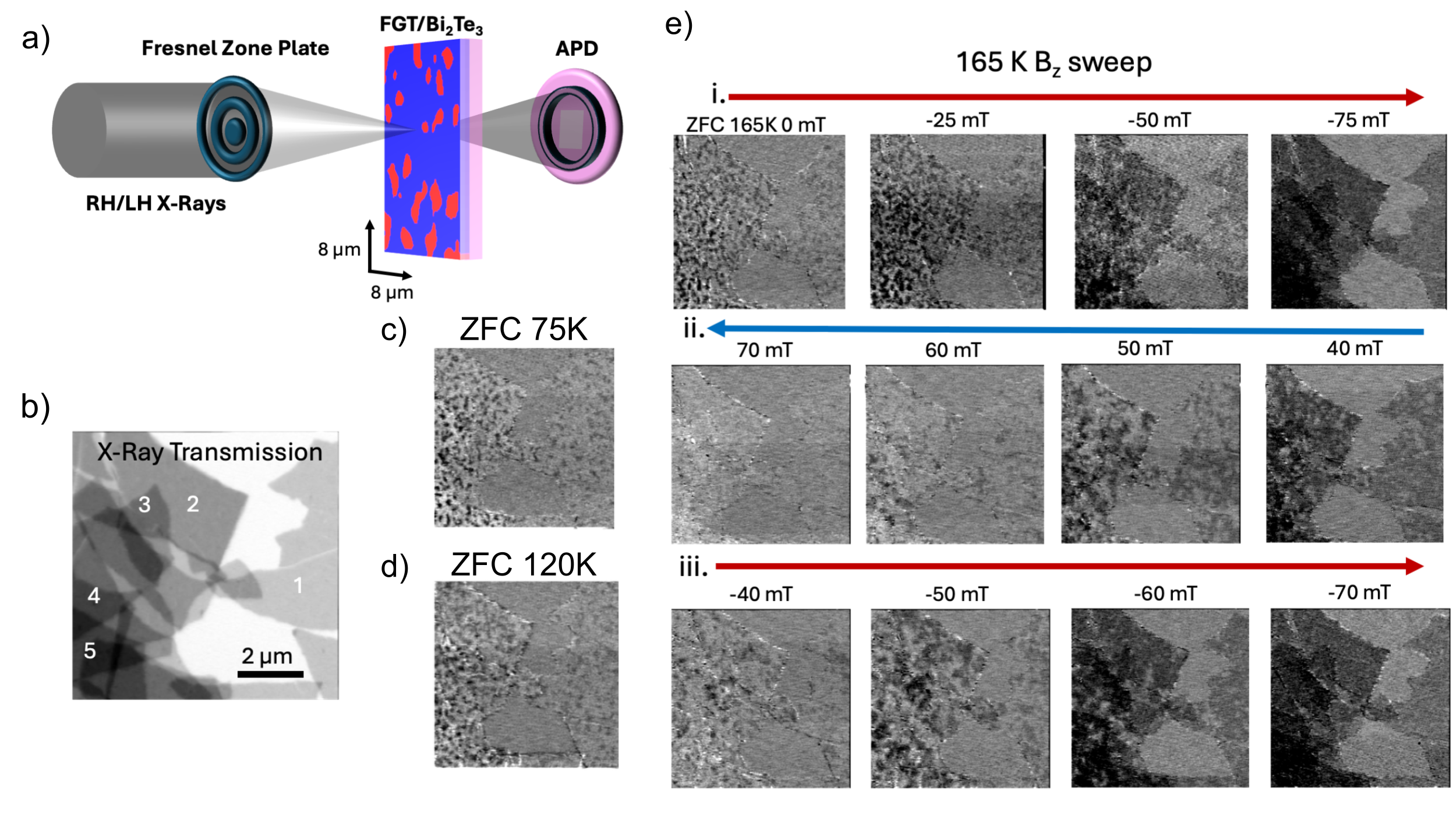}
    \caption{\textbf{X-ray magnetic circular dichroism images of bubble domains stabilized in an epitaxial Fe$_3$GeTe$_2$/Bi$_2$Te$_3$ heterostructure transferred onto a SiRN membrane.} (a) Schematic of scanning transmission X-ray magnetic circular dichroism measurements of nanoscale bubble domains. All XMCD images were acquired over an 8 $\mu$m $\times$ 8 $\mu$m raster scan with a step size of 40 nm. (b) X-ray transmission averaged between right and left circularly polarized X-rays, of regions with different number of heterostructure repeats [FGT/ Bi$_2$Te$_3$]$_n$ (n = 1 to 5) from folding during the transfer process. (c-d) XMCD images of the zero-field-cooled (ZFC) state at 75 K and 120 K. e) XMCD images at different magnetic fields taken sequentially during the out-of-plane field hysteresis sweep directions: (i) negative, (ii) positive, (iii) negative following the zero-field cooling to 165 K.}
    \label{Figure2}
\end{figure}

To directly visualize the magnetic domain morphology in epitaxial FGT/Bi$_2$Te$_3$ heterostructures, we performed circularly polarized scanning transmission X-ray microscopy (XMCD-STXM) at the Fe $L_3$ absorption edge (709 eV), determined by measuring the absorption spectra (S.I.IV). XMCD contrast is obtained from the normalized difference between STXM images acquired with opposite circular polarizations, providing an element-specific quanitative probe of the local out-of-plane magnetization component,  M$_z$. The measurement principle is illustrated schematically in Fig.~\ref{Figure2}a. All XMCD-STXM measurements were conducted at the MAXYMUS endstation at the BESSY II synchrotron (see methods). We performed the measurements at three different temperatures below T$_c$: 75 K, 120 K and 165 K. At each temperature, the sample was cooled from 300 K without any external field (zero-field-cooled, ZFC). XMCD images were first acquired in the ZFC state, before any fields were applied. This was then followed by sequential measurements during an out-of-plane magnetic-field hysteresis sweep.  

The TRT transfer process produces folds in the FGT/Bi$_2$Te$_3$ film, resulting in regions with different number of [FGT/Bi$_2$Te$_3$]$_n$ layer repeats (n = 1 to 5) corresponding to different total FGT thicknesses (t$_{FGT}$ = 6.4 nm to 32 nm) \cite{li2025fullfilmdrytransfermbegrown}. The number of these heterostructure repeats are readily identified by variations in the STXM transmission amplitude contrast (Fig.~\ref{Figure2}b). Under ZFC conditions, XMCD images consistently reveal dark, circular features on a bright background across all thicknesses and in all three temperatures studied. These features correspond to magnetic bubble domains with magnetization oriented antiparallel to the film normal (-M$_z$). As shown in Fig.~\ref{Figure2}c-e, these domain textures remain remarkably similar across all thicknesses measured, with no discernible changes in domain size or areal density. The XMCD contrast systematically increases with effective thickness as shown in Fig.~\ref{Figure3}b, consistent with vertically extended, cylindrical bubble columns spanning the full film thickness. This behavior indicates ferromagnetic coupling between FGT layers across the [FGT/Bi$_2$Te$_3$]$_n$ superlattice, likely mediated by long-range dipolar interactions. Notably, the persistence of nearly identical domain morphologies across all thicknesses demonstrates that repeated stacking of FGT/Bi$_2$Te$_3$ does not alter the equilibrium domain texture; instead, the magnetic morphology is dictated primarily by interfacial interactions with Bi$_2$Te$_3$ within each individual fold.

\vspace{0.5cm}
Our XMCD-STXM results differ from prior studies of intrinsic, exfoliated FGT flakes in two key respects. First, magnetic bubble domains in the FGT/Bi$_2$Te$_3$ heterostructures emerge robustly under zero-field-cooled (ZFC) conditions across all heterostructure repeats, whereas exfoliated FGT flakes typically exhibit stripe-domain ground states after ZFC, with bubble or skyrmion textures appearing only under applied magnetic fields or following field cooling \cite{Birch2022-oh, Birch_2024, Garland2025}. Second, the bubble-domain phase in our heterostructures exhibit similar characteristics over a wide effective thickness range, from 6.4 to 32 nm of FGT (corresponding to n = 1 to 5 repeats). This is distinct from previous literature measurements on thin exfoliated flakes $<$ 15 nm that only report mono-domain characteristics \cite{Fei2018} or in-plane anisotropy \cite{WangHong2020}, while the equilibrium domain type, characteristic domain density and dimensions vary strongly with thickness for thicker flakes \cite{Birch2022-oh, Chakraborty2022, Garland2025}. Hence, to our knowledge, our results constitute the first observation of ZFC bubble domains in thin FGT layers (6.4 nm). Furthermore, the presence of domains with similar characteristic sizes and densities across temperatures and heterostructure repeats (n = 1 to 5) under ZFC conditions suggests that proximity to Bi$_2$Te$_3$ plays a central role in enabling interfacial stabilization mechanisms absent in exfoliated FGT flakes. In our epitaxial heterostructures, this interfacial mechanism dominates domain morphology over thickness, temperature and magnetic history, providing a new route to stabilizing target spin textures with arbitrary total magnetic volume using vdW topological insulator/magnet superlattices.

\vspace{0.5cm}
We next examine the evolution of the bubble domains under a uniform out-of-plane external magnetic field with varying magnitude. As shown in Fig.~\ref{Figure2}f-h, at 165 K and starting from the ZFC state, increasing the applied field in the negative direction leads to a gradual growth of bubble domains between 0 and -50 mT, accompanied by coalescence into larger patch-like regions. At -75 mT, the film reaches a uniformly saturated state with magnetization aligned along the -M$_z$ direction. Upon sweeping the field toward positive values, nonuniform domain states re-emerge at +40 mT, +50 mT, and +60 mT. Specifically, at +40 mT, white bubble domains nucleate on a dark XMCD background and progressively expand into larger white patches as the field increases to +50 mT (Fig.~\ref{Figure2}g). At +60 mT, these patches further grow and merge, forming a near-uniform XMCD contrast, followed by full saturation along the +M$_z$ direction at +70 mT. The subsequent field sweep back toward negative saturation exhibits analogous behavior, with dark bubble domains nucleating on a bright background near -40 mT and expanding continuously with increasing negative field (Fig.~\ref{Figure2}h). Notably, throughout the entire out-of-plane field hysteresis loop, no transition to stripe-domain phases is observed at any field value or temperatures studied; data at 75 K and 120 K show similar behavior (SI.III). These results further confirm that bubble domains constitute the dominant nonuniform magnetic texture in the epitaxial FGT/Bi$_2$Te$_3$ heterostructures.

\subsection*{Magnetic Domain Analysis}
\begin{figure}[htpb]
    \centering
    \includegraphics[width=\textwidth]{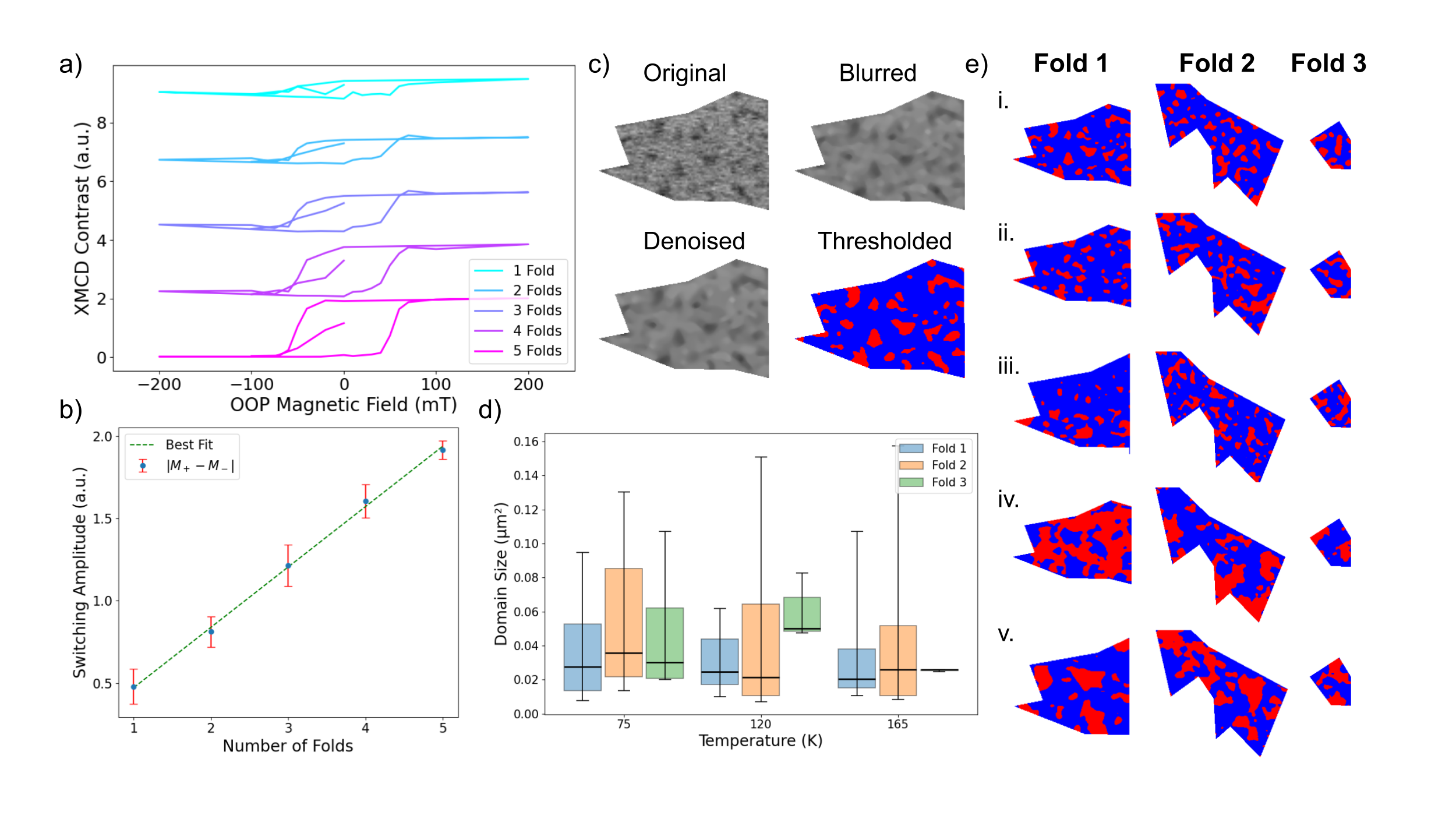}
    \caption{\textbf{Characterization of magnetization and magnetic domains based on XMCD-STXM contrast imaging.} The spatially averaged XMCD grayscale value is normalized between 0 and 1 and used as the effective magnetization. (a) Magnetic field sweep hysteresis for regions with [FGT/Bi$_2$Te$_3$]$_n$ heterostructure repeats from n = 1 to 5 at 165 K. Hysteresis loops of each region is scaled and artificially vertically offset for clarity. (b) Saturated XMCD switching amplitudes as a function of repeats, exhibiting linear scaling with effective magnetic thickness. (c) Representative image-processing pipeline for domain identification, including median blurring, a non-local means denoising, and threshold-based segmentation (S.I.VII; example shown for a one-fold region at 75 K). (d) Zero-field domain-size distributions extracted from the processed XMCD for fold 1-3 at 75, 120, and 160 K; box plots indicating the 5th, 25th, 50th, 75th, and 95th percentiles. (e) Thresholded images showing bubble domains in fold 1-3: (i-iii) zero-field states at 75, 120, and 165 K, and (iv-v) field-driven states at 165 K under 50 mT and –50 mT, respectively.}
    \label{Figure3}
\end{figure}

We next quantitatively analyze the XMCD contrast associated with bubble domains as a function of temperature and effective magnetic thickness. Regions with different heterostructure repeats were analyzed separately, and the spatially averaged XMCD contrast was extracted as a function of the hysterically applied out-of-plane magnetic field (Fig.~\ref{Figure3}a,b; S.I.V and S.I.VI). As shown in Fig.~\ref{Figure3}a, all five thickness regions exhibit nearly identical hysteresis loop shapes with comparable coercive fields, closely matching the optical MCD response of the as-grown films (Fig.~\ref{Figure1}c). This agreement confirms that neither the TRT transfer process nor X-ray exposure degrades the magnetic properties. We further extract the switching magnitude of spatially averaged XMCD contrast between the +M$_z$ and -M$_z$ saturated states for each fold number. While the resulting switching amplitude increases linearly with effective thickness (Fig.~\ref{Figure3}b), the coercive field and overall hysteresis softness remain largely unchanged, reinforcing that the observed nonuniform magnetic domain texture are largely independent of effective thickness.

\vspace{0.5cm}
To quantitatively delineate and classify individual bubble domains, we implement a standardized image-processing pipeline \cite{OpenCV} to the XMCD-STXM images (Fig.~\ref{Figure3}c, S.I.VII). This procedure enables reliable identification of individual domains, which are then used to construct domain-size distributions as a function of temperature and effective magnetic thickness. As summarized in Fig.~\ref{Figure3}d, we observe no systematic dependence of the domain-size distribution on either the number of heterostructure repeats or temperature (detailed statistical analysis in S.I.VIII). Across all conditions studied, the characteristic bubble diameter remains on the order of 100 nm, consistent with previously reported bubble and skyrmion phases in exfoliated FGT \cite{Wu2020-yy, Wu2022-yy, Birch2022-oh}. Representative thresholded XMCD images for one to three heterostructure repeats at different temperatures and field conditions are shown in Fig.~\ref{Figure3}e, providing direct visual confirmation of the uniformity and robustness of the bubble domain morphology. In the following sections, we combine first-principles calculations and micromagnetic simulations to elucidate how interfacial interactions with Bi$_2$Te$_3$, including proximity-modified magnetic anisotropy and interfacial Dzyaloshinskii–Moriya interaction, stabilize this bubble-domain phase.

\subsection*{First-Principle Calculation and Micromagnetic Simulations}
\begin{figure}[htpb]
    \centering
    \includegraphics[width=\textwidth]{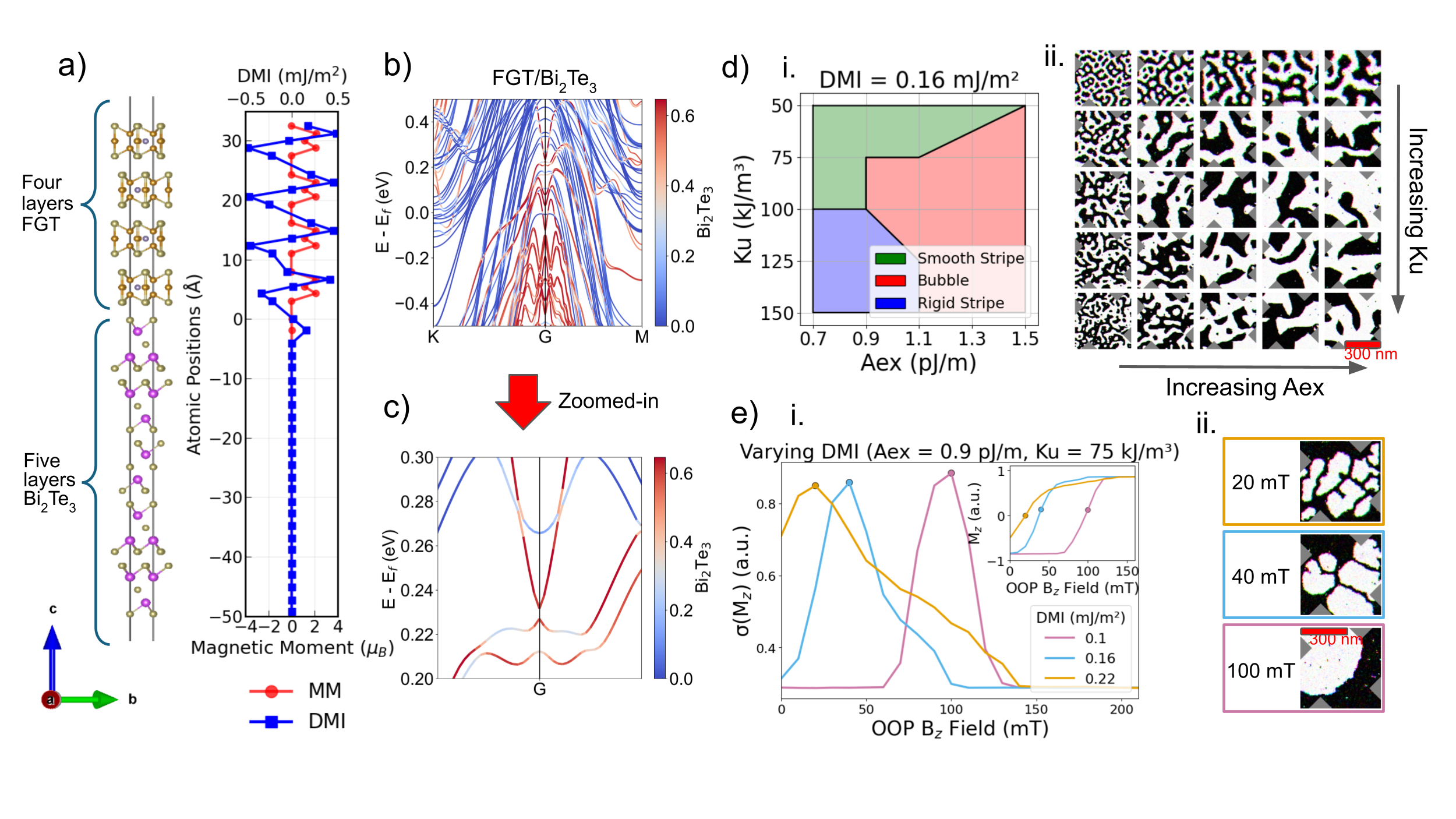}
    \caption{\textbf{First-principles calculations and micromagnetic simulations of bubble-domain stabilization.} 
   (a) Slab supercell used in the DFT calculations together with the corresponding atomic-layer--resolved magnetic moment (MM, red) and Dzyaloshinskii--Moriya interaction (DMI, blue), aligned along the out-of-plane direction. Bi, Te, Fe, and Ge atoms are shown in magenta, brass, brown, and purple-blue, respectively; the topmost Te layer of Bi$_2$Te$_3$ defines the zero reference. (b, c) Projected band structure of the heterostructure and a zoomed-in view near the Fermi level, showing Fe$_3$GeTe$_2$-derived (blue) and Bi$_2$Te$_3$-derived (red) states. (d) Zero-field domain morphologies across micromagnetic parameter space with DMI = 0.16 mJ/m$^2$, exchange stiffness (Aex) = 0.7–1.5 pJ/m, and uniaxial anisotropy (Ku) = 50–150 kJ/m$^3$ (i) Domain classifications: smooth stripe (green), bubble (peach), and rigid stripe (blue). (ii) Corresponding simulated domain snapshots. (e) Field-induced switching simulations for Aex = 0.9 pJ/m, Ku = 75 kJ/m$^3$, and Dind = 0.10 (purple), 0.16 (teal), and 0.22 (orange) mJ/m$^2$. (i) Evolution of the standard deviation of out-of-plane magnetization ($\sigma(M_z)$) as a function of out-of-plane (OOP) magnetic field ($B_z=0$--$200$~mT). The inset shows $M_z$  versus $B_z$. Filled symbols mark the field of maximum nucleation at $B_z$ = 100, 40, and 20 mT for DMI = 0.10, 0.16, and 0.22 mJ/m$^2$, respectively (see details in Fig.~S15). (ii) Corresponding domain snapshots at maximum nucleation.}
    \label{Figure4}
\end{figure} 
To elucidate the microscopic origin of the bubble-domain stabilization in FGT/Bi$_2$Te$_3$, we performed first-principles calculations based on a slab geometry comprising Fe$_3$GeTe$_2$ interfaced with Bi$_2$Te$_3$. Layer-resolved magnetic moment analysis shows that the magnetization is strongly localized on the Fe atoms within the Fe$_3$GeTe$_2$ layers, with no appreciable magnetic moment induced on Bi or Te atoms in Bi$_2$Te$_3$ (Fig.~\ref{Figure4}a red curve), consistent with the nonmagnetic bulk character of Bi$_2$Te$_3$. In contrast, the layer-resolved Dzyaloshinskii--Moriya interaction (DMI) exhibits a distinct interfacial character (Fig.~\ref{Figure4}a blue curve). Within the Fe$_3$GeTe$_2$ layers, the DMI contributions alternate in sign from layer to layer due to the frustrated triangular lattice of Fe atoms, leading to partial cancellation in the absence of inversion symmetry breaking. At the FGT/Bi$_2$Te$_3$ interface, inversion symmetry is broken, resulting in a finite interfacial DMI localized near the interface that does not cancel. A small but finite DMI contribution is also induced on Bi atoms in the interfacial Bi$_2$Te$_3$ layer despite the absence of an induced magnetic moment, highlighting the role of strong spin--orbit coupling rather than magnetic proximity effects. Spin-spiral calculations yield an interfacial Dzyaloshinskii--Moriya interaction strength of $D \approx -1.5 \times 10^{-4}$~J\,m$^{-2}$, which is sufficient to stabilize chiral bubble domains in the presence of perpendicular magnetic anisotropy (S.I.X).

\vspace{0.5cm}
The electronic origin of this interfacial interaction is further reflected in the projected band structure of the FGT/Bi$_2$Te$_3$ heterostructure (Fig.~\ref{Figure4}b,c). Near the Fermi level, Fe$_3$GeTe$_2$-derived states hybridize with Bi$_2$Te$_3$-derived states, while the topological surface state of the Bi$_2$Te$_3$ slab remains intact but is shifted above the Fermi level. This interfacial hybridization provides a microscopic basis for enhanced spin--orbit coupling at the interface, which under broken inversion symmetry gives rise to a finite interfacial DMI.

\vspace{0.5cm}
To translate the interfacial interactions identified by DFT into real-space magnetic textures, we perform micromagnetic simulations that systematically explore how exchange stiffness (Aex), uniaxial magnetic anisotropy (Ku), and interfacial DMI jointly determine the equilibrium domain morphology. Figure~\ref{Figure4}d summarizes resulting zero-field domain morphologies across a broad micromagnetic parameter space at fixed interfacial DMI, revealing three distinct
nonuniform textures: smooth stripe domains, bubble domains, and rigid stripe domains. Bubble domains occupy an intermediate region of parameter space where exchange, anisotropy, and interfacial DMI are comparably balanced, while stripe-like phases dominate when either exchange or anisotropy strongly outweighs the chiral interaction. Within this regime, domain walls are highly curved and form mostly closed, bubble structures whose characteristic sizes are comparable to the experimentally observed domains in Fig.~\ref{Figure3}d (Fig.~S14), whereas stripe-dominated phases are characterized by more extended domains with straighter walls. The phase boundaries evolve continuously, indicating crossover behavior rather than sharp phase transitions, a trend that persists across a broader range of interfacial DMI strengths (Fig.~S13).
\vspace{0.5cm}

The field-driven evolution of these domain textures is captured by micromagnetic simulations under out-of-plane magnetic-field sweeps, as summarized in Fig.~\ref{Figure4}e. Starting from the zero-field bubble-domain state, increasing field progressively suppresses nonuniform magnetization and drives the system toward saturation, while field reversal induces domain nucleation and growth over a finite field interval (S.I.X). This process is quantitatively tracked using the standard deviation of the out-of-plane magnetization, $\sigma(M_z)$, which exhibits a pronounced peak during field reversal corresponding to maximal domain formation, as shown in Fig.~\ref{Figure4}e(i), the position and width of this $\sigma(M_z)$ peak depend sensitively on the interfacial DMI strength, reflecting changes in the field range over which bubble domains nucleate and persist. The corresponding domain snapshots (Fig.~\ref{Figure4}e(ii)) confirm that the maximum in $\sigma(M_z)$ coincides with dense bubble-domain configurations, whereas low $\sigma(M_z)$ values correspond to nearly uniform magnetization. In S.I.X, the dependence of the hysteresis loops on exchange stiffness, anisotropy, and DMI is examined, the use of $\sigma(M_z)$ as a quantitative proxy for domain nucleation is detailed, and extended parameter-dependent field-sweep analyses together with the corresponding domain snapshots are provided. These results show that the interfacial interactions responsible for zero-field bubble stability also control their field-driven evolution, consistent with experiment.

\subsection*{Summary and Outlook}

In summary, we have demonstrated direct nanoscale imaging of magnetic domain textures in MBE-grown epitaxial Fe$_3$GeTe$_2$/Bi$_2$Te$_3$ heterostructures and uncovered a robust bubble-domain regime stabilized under zero-field-cooled conditions across a wide range of temperatures and effective thicknesses. XMCD-STXM measurements combined with quantitative domain analysis reveal that the bubble-domain morphology is remarkably insensitive to effective thickness variations, temperature, and magnetic-field history over the experimentally accessible range, distinguishing this epitaxial heterostructure platform from exfoliated Fe$_3$GeTe$_2$ flakes, where the type of domain (stripe, bubble skyrmion) is strongly thickness, temperature and field-history-dependent. By integrating first-principles calculations with micromagnetic simulations, we identify the microscopic origin of this behavior. Proximity to Bi$_2$Te$_3$ breaks inversion symmetry and introduces strong interfacial spin–orbit coupling, generating a finite interfacial Dzyaloshinskii–Moriya interaction and renormalizing the effective magnetic anisotropy governing domain formation in Fe$_3$GeTe$_2$. These interfacial modifications reshape the magnetic phase landscape, placing the system in a regime where bubble domains are energetically favored at zero field and exhibit characteristic sizes and field responses consistent with experimental observations.

\vspace{0.5cm}
More broadly, our results highlight interfacial engineering as a powerful and flexible route to controlling magnetic domain phases in atomically thin vdW magnets without relying on geometric confinement, field cooling, or thickness control. In these vdW heterostructures, interfacial spin–orbit coupling, DMI, and effective anisotropy can be further tuned through the choice of topological insulator, interface termination, strain, or electrostatic gating, providing a versatile pathway toward controllable bubble domains, skyrmions, and related chiral textures with arbitrary total magnetic volumes. Extending this framework to other magnetic materials and spin--orbit--coupled substrates may enable systematic design of target domain phases, while time-resolved, current-driven, and electrically or optically gated studies may offer promising routes to elucidate and control the dynamics of these interfacially stabilized textures. Together, these opportunities position vdW topological-insulator/magnet heterostructure superlattices as a powerful platform for both fundamental studies of interfacial magnetism and the development of reconfigurable spintronic and magnonic functionalities.

\newpage
\bibliography{scibib}
\bibliographystyle{Science}

\section*{Methods}
The high quality FGT/Bi$_2$Te$_3$ films on Al$_2$O$_3$ substrates are grown in a Veeco 930 MBE system. Bi$_2$Te$_3$ layers are grown by codepositing Bi and Te in two steps successively (265$\degree$C and 300$\degree$C), followed by codepositing Fe, Ge and Te at 325$\degree$C for growths of FGT layers. Te flux is maintained all the time to achieve Te overpressure and prevent any Te deficiency in the atomic structures. The thermal release tape transfer (TRT) process was then performed as follows: a 5\% solution (by weight) of polycaprolactone (PCL, Sigma Aldrich, average M$_n$: 80000) in tetrahydrofuran (THF, Sigma Aldrich, $\geq99.9\%$ anhydrous) is spin coated at 1000 rpm for 1 min on top of the as-grown Fe$_3$GeTe$_2$/Bi$_2$Te$_3$, followed with a 5-min bake at 75$\degree$C to form a uniform PCL film. A piece of thermal release tape (TRT, Nitto Denko Corp, Revalpha RA-95LS(N)) and a polydimethylsiloxane (PDMS, Gel-Pak) stamp are gently attached to the PCL film for mechanical support. The entire structure is then slowly peeled off by tweezer and then aligned and approached to the target silicon-rich ntride membrane substrate by a home-built transfer tool. The PCL film is then fully melted at 90$\degree$C and then dissolved in THF solvent at room temperature.
Silicon-rich nitride membranes were purchased from Silson Ltd. with a  stoichiometry of approximately 1:1 and a lower density and stress than that of Si$_3$N$_4$, allowing for a higher transmission. The silicon frame is 5.0 mm x 5.0 mm and 200 $\mu$m thick, with the membrane being 0.25 mm x 0.25 mm and 100 nm thick. MCD measurements were performed at The Ohio State University. Further details in S.I.I. and S.I.II.

X-ray magnetic circular dichroism based on scanning transmission X-ray microscopy (XMCD-STXM) measurements were performed at the Magnetic X-ray Microscope with UHV Spectroscopy (MAXYMUS) instrument on the UE46 beamline at the BESSY II electron storage ring operated by the Helmholtz-Zentrum Berlin für Materialien und Energie. Samples were mounted inside a closed system microscope and cooled with a He cryostat, with any applied magnetic field up to 250 mT determined by adjusting the orientation of four permanent magnets near the sample space. The X-ray resolution limit was 20 nm, resulting from focusing from a Fresnel zone plate and order selection aperture (OSA), rastering across the sample with a piezoelectric sample stage, and subsequent detection by an avalanche photodiode detector (APD). Out of plane magnetic contrast was determined by exploiting the X-ray resonance energy at the 709 eV Fe-L3 edge. Image processing was achieved through subtracting two polarizations, one positive circular polarization and the other negative circular polarization, revealing a pure XMCD magnetic contrast. 

XMCD analysis was performed using an OpenCV-based image-processing pipeline \cite{OpenCV}. Thickness-dependent masks were traced from the X-ray transmission image (Fig.\ref{Figure3}b) and propagated across the measurement sequence via drift correction (S.I.VII). Within each mask, filters \texttt{cv2.medianBlur()} and \texttt{cv2.fastNlMeansDenoising()} were applied to suppress grid-like distortions in the XMCD images (S.I.VII and VIII). Magnetic domains were isolated using \texttt{cv2.threshold()}, with thresholds selected individually by comparison of the original, filtered, and thresholded images (S.I.IX). Domain statistics were extracted using \texttt{cv2.connectedComponentsWithStats()}, which identifies and characterizes individual connected regions corresponding to magnetic bubble domains.

First-principle calculations were performed using VASP \cite{KressePRB1993,KressePRB1994,KresseCompMatSci1996,KressePRB1996} and FLEUR \cite{fleurWeb} software packages within the local-density approximation (LDA) for exchange correlation\cite{SahniPRA1988}. We selected the heterostructure with an in-plane lattice constant of a = 4.00 \AA for further analysis, as its out-of-plane magnetic anisotropy is consistent with experimental observations. Structural relaxation was carried out by allowing atomic positions to relax while keeping the cell shape and volume fixed. The calculations employed a 7×7×1 k-mesh and an energy cutoff of 415 eV. The relaxation criterion was set to EDIFF = 10$^{-4}$, resulting in average and maximum residual forces of 0.040 eV/\AA and 0.149 eV/\AA, respectively, upon convergence. For FLEUR, we used a plane-wave cutoff of 4.0 a.u-1 and k-point mesh of 9x9x1 in thin-film geometry. Spin spiral calculations were made within the generalized Bloch theorem formalism. The exchange stiffness and DMI strength were obtained using magnetic force theorem within first-perturbation theory, as implemented in FLEUR. Magnetocrystalline anisotropic energies were calculated using second-variation of spin-orbit coupling scheme.
\vspace{0.5cm}

\noindent \textbf{Acknowledgements:} We thank Roberto Panepucci and Xiaoxi Huang for experimental assistance and discussions. We acknowledge experimental assistance from the MAXYMUS UE46-PGM2 beamline at BESSY II synchrotron in Helmholtz-Zentrum Berlin and the Cornell NanoScale Facility (NSF Grant NNCI-2025233).

\vspace{0.5cm}
\noindent \textbf{Funding:} The research at USC is supported by DOE BES program award number DE-SC0025422, ARO ECP contract
number W911NF2510276. The research at Cornell was supported by the AFOSR/MURI project 2DMagic (FA9550-19-1-0390) and the US National Science Foundation through DMR-2104268 and the Cornell Center for Materials Research (DMR-1719875). T.M.J.C. was supported in part by the Caltech David and Ellen Lee Postdoctoral Research Scholarship. M. Z. acknowlege University of Southern California PURF and SURF funds. Y.K.L. acknowledges a Cornell Presidential Postdoctoral Fellowship.

\vspace{0.5cm}
\noindent \textbf{Author contributions:} W.Z. performed the epitaxial growth of the vdW films and Kerr rotation characterization of the films supervised by R.K.K.. W.Z. and Z.L. performed the thermal release transfer of the films onto STXM-compatible SiRN membranes. C.S. designed and performed the magnetic force microscopy measurements. S.A. performed focused ion beam milling supervised by J.W.. T.M.J.C, Y.K.L., S.W. and L.P. performed the XMCD measurements with assistance from A.K., D.B. and M.W.. D-K.L. and E.T. performed first-principles calculations supervised by J.G.. M.Z. performed the micromagnetic simulations with assistance from T.M.J.C. supervised by Y.K.L.. R.K.K. provided oversight and advice. T.M.J.C., M.Z., A.K. and Y.K.L. wrote the manuscript, with help from the other authors. All authors discussed the results and gave feedback on the manuscript.

\vspace{0.5cm}
\noindent \textbf{Competing interests:} The authors declare no competing interests.

\vspace{0.5cm}
\noindent \textbf{Data and materials availability:} All data from the main text and supplementary materials will be deposited in the Zenodo repository prior to publication, and a link will be provided upon acceptance of the paper.

\end{document}